\documentclass[12pt,epsf]{article}
\usepackage{graphicx,amsmath,amssymb}
\begin{document}
%%%%%%%%%%%%%%%%%%%%%%%%%%%%%%%%%%%%%%%%%%%

\def\a{\alpha}
\def\b{\beta}
\def\c{\varepsilon}
\def\d{\delta}
\def\e{\epsilon}
\def\f{\phi}
\def\g{\gamma}
\def\h{\theta}
\def\k{\kappa}
\def\l{\lambda}
\def\m{\mu}
\def\n{\nu}
\def\p{\psi}
\def\q{\partial}
\def\r{\rho}
\def\s{\sigma}
\def\t{\tau}
\def\u{\upsilon}
\def\v{\varphi}
\def\w{\omega}
\def\x{\xi}
\def\y{\eta}
\def\z{\zeta}
\def\D{\Delta}
\def\G{\Gamma}
\def\H{\Theta}
\def\L{\Lambda}
\def\F{\Phi}
\def\P{\Psi}
\def\S{\Sigma}

\def\o{\over}
\def\beq{\begin{eqnarray}}
\def\eeq{\end{eqnarray}}
\newcommand{\gsim}{ \mathop{}_{\textstyle \sim}^{\textstyle >} }
\newcommand{\lsim}{ \mathop{}_{\textstyle \sim}^{\textstyle <} }
\newcommand{\vev}[1]{ \left\langle {#1} \right\rangle }
\newcommand{\bra}[1]{ \langle {#1} | }
\newcommand{\ket}[1]{ | {#1} \rangle }
\newcommand{\EV}{ {\rm eV} }
\newcommand{\KEV}{ {\rm keV} }
\newcommand{\MEV}{ {\rm MeV} }
\newcommand{\GEV}{ {\rm GeV} }
\newcommand{\TEV}{ {\rm TeV} }
\def\diag{\mathop{\rm diag}\nolimits}
\def\Spin{\mathop{\rm Spin}}
\def\SO{\mathop{\rm SO}}
\def\O{\mathop{\rm O}}
\def\SU{\mathop{\rm SU}}
\def\U{\mathop{\rm U}}
\def\Sp{\mathop{\rm Sp}}
\def\SL{\mathop{\rm SL}}
\def\tr{\mathop{\rm tr}}

\def\IJMP{Int.~J.~Mod.~Phys. }
\def\MPL{Mod.~Phys.~Lett. }
\def\NP{Nucl.~Phys. }
\def\PL{Phys.~Lett. }
\def\PR{Phys.~Rev. }
\def\PRL{Phys.~Rev.~Lett. }
\def\PTP{Prog.~Theor.~Phys. }
\def\ZP{Z.~Phys. }

%%%%%%% added by Fumi %%%%%%%%%%
% FROM HERE
%\newcommand{\beq}{\begin{equation}}   
%\newcommand{\eeq}{\end{equation}}
\newcommand{\bea}{\begin{eqnarray}}   
\newcommand{\eea}{\end{eqnarray}}
\newcommand{\bear}{\begin{array}}  
\newcommand {\eear}{\end{array}}
\newcommand{\bef}{\begin{figure}}  
\newcommand {\eef}{\end{figure}}
\newcommand{\bec}{\begin{center}}  
\newcommand {\eec}{\end{center}}
\newcommand{\non}{\nonumber}  
\newcommand {\eqn}[1]{\beq {#1}\eeq}
\newcommand{\la}{\left\langle}  
\newcommand{\ra}{\right\rangle}
\newcommand{\ds}{\displaystyle}
\def\SEC#1{Sec.~\ref{#1}}
\def\FIG#1{Fig.~\ref{#1}}
\def\EQ#1{Eq.~(\ref{#1})}
\def\EQS#1{Eqs.~(\ref{#1})}
\def\GEV#1{10^{#1}{\rm\,GeV}}
\def\MEV#1{10^{#1}{\rm\,MeV}}
\def\KEV#1{10^{#1}{\rm\,keV}}
\def\lrf#1#2{ \left(\frac{#1}{#2}\right)}
\def\lrfp#1#2#3{ \left(\frac{#1}{#2} \right)^{#3}}
% UNTIL HERE

%%%%%%%%%%%%%%%%%%%%%%%%%%%%%%%%%%%%%%%%%%%%%%%%%%%%%%%%%%%%%%%%%%%%

\baselineskip 0.7cm

\begin{titlepage}

\begin{flushright}
UT-13-21\\
TU-937\\
IPMU13-0101\\
\end{flushright}

\vskip 1.35cm
\begin{center}
{\large \bf 
Polynomial Chaotic Inflation in Supergravity
}
\vskip 1.2cm
Kazunori Nakayama$^{a,c}$,
Fuminobu Takahashi$^{b,c}$
and 
Tsutomu T. Yanagida$^{c}$

\vskip 0.4cm

{\it $^a$Department of Physics, University of Tokyo, Tokyo 113-0033, Japan}\\
{\it $^b$Department of Physics, Tohoku University, Sendai 980-8578, Japan}\\
{\it $^c$Kavli Institute for the Physics and Mathematics of the Universe (WPI), TODIAS, University of Tokyo, Kashiwa 277-8583, Japan}

\vskip 1.5cm

\abstract{

We present a general polynomial chaotic inflation model in supergravity, 
for which the predicted spectral index and tensor-to-scalar ratio can lie
within the $1 \sigma$ region allowed by the Planck results. Most importantly,
the predicted tensor-to-scalar ratio is large enough to be probed in the on-going and 
future B-mode experiments. We study the inflaton dynamics and the subsequent 
reheating process in a couple of specific examples. The non-thermal gravitino production
from the inflaton decay can be suppressed in a case with a discrete $Z_2$ symmetry.
We find that the reheating temperature can be naturally
as high as $O(10^{9-10})$\,GeV, sufficient for baryon asymmetry generation through (non-)thermal 
leptogenesis. 
}
\end{center}
\end{titlepage}

\setcounter{page}{2}

%%%%%%%%%%%%%%%%%%%%%%%%%%%%%%%%%%%%%
\section{Introduction}
%%%%%%%%%%%%%%%%%%%%%%%%%%%%%%%%%%%%%
How did the Universe begin? That is one of the most important questions in
cosmology and particle physics. The fact that the observed cosmic microwave
background (CMB)  has superhorizon-sized correlations  implies that our Universe experienced a stage of accelerated expansion, i.e. 
inflation~\cite{Guth:1980zm,Linde:1981mu},  at a very early stage of evolution.
Recently, the Planck satellite~\cite{Ade:2013rta} measured the CMB temperature anisotropies 
with unprecedented precision, showing that the  CMB power spectrum
can be well fitted by nearly scale-invariant, adiabatic and Gaussian 
density perturbations. This gives  strong preference to a canonical (effectively) single-field inflation.

Among various inflation models proposed so 
far~\cite{Martin:2013tda,Yamaguchi:2011kg}, a chaotic inflation model~\cite{Linde:1983gd} is particularly interesting, not only because it avoids the fine tuning of the initial condition for inflation, but also because it
predicts a large tensor-to-scalar ratio that is within the reach of the Planck satellite
and future CMB observation experiments. The primordial tensor mode, if detected, will provide us with the absolute
energy scale of inflation, and we can get invaluable information on the very early Universe. 

The chaotic inflation requires a super-Planckian value for the inflaton field, which was
an obstacle for model building in supergravity (SUGRA) and superstring theories. This problem was surmounted
in a simple model proposed in Ref.~\cite{Kawasaki:2000yn}, where
the K\"ahler potential of the following form was introduced,
\begin{equation}
	K = \frac{1}{2}(\phi + \phi^\dagger)^2,
\end{equation}
 which respects a shift symmetry:
\begin{equation}
	\phi \to \phi + i \alpha
	\label{shift}
\end{equation}
with $\alpha$ being a real constant. Here and in what follows we adopt the Planck units where
the reduced Planck mass $M_P \simeq 2.4 \times \GEV{18}$ is set to be unity.
Then the inflaton,  $\varphi \equiv \sqrt{2}\, {\rm Im}\phi$,
 can take a super-Planckian field value without receiving 
any dangerous SUGRA corrections.
Combined with a superpotential $W = mX\phi$ with $X$ being another chiral superfield and $m$ the inflaton mass, 
$\varphi$ has a simple quadratic potential beyond the Planck scale, leading to chaotic 
inflation. Note that, in this model as well as the inflation models along the lines of Ref.~\cite{Kawasaki:2000yn},
 the inflaton $\varphi$ is a real scalar field, not a complex one, and therefore there
is no degrees of freedom that may destabilize the inflationary path.
Up to now, there are various models of chaotic inflation developed in 
supergravity~\cite{Kawasaki:2000ws, Kallosh:2010ug,Kallosh:2010xz,Takahashi:2010ky,Nakayama:2010kt,Harigaya:2012pg,Nakayama:2013jka} and string theory~\cite{Silverstein:2008sg,McAllister:2008hb,Peiris:2013opa}.

Much attention has been paid to a simple class of chaotic inflation models 
based on a monomial potential, $V(\varphi) \propto \varphi^n$. 
Some of the models in this class are already in tension
with the Planck observation of the spectral index $(n_s)$ and the tensor-to-scalar ratio $(r)$.
Specifically, a chaotic inflation model based on a quartic potential is  highly disfavored because of
too large $r$,  and that  based on a quadratic potential is marginally consistent with the observation
at $2 \sigma$ level. Those with a linear or fractional power potential lie
outside $1\sigma$ but within $2\sigma$ allowed region. It is worth noting that  no model in this class
 lies within $1\sigma$ allowed region.

Motivated by the possible tension of the monomial chaotic inflation with the observation,
the present authors have recently proposed a polynomial chaotic inflation 
model in SUGRA~\cite{Nakayama:2013jka}, where a superpotential of the following form was introduced,
\begin{equation}
	W = X(m\phi + \lambda \phi^2),
	\label{W2}
\end{equation}
with $\lambda$ being a constant parameter.
Surprisingly, this simple (and possibly natural) extension can cover almost all the values of
 $n_s$ and $r$ within the 1$\sigma$ region of the Planck result~\cite{Nakayama:2013jka}.
The point is that the addition of $\phi^2$-term can make the scalar potential  flatter at $\varphi \gtrsim 1$
and, as a result, the tenor-to-scalar ratio can be reduced while $n_s$ remains within the $1\sigma$ allowed region.

Lastly we briefly mention related works. In Ref.~\cite{Destri:2007pv}, the inflation model based on a scalar 
potential equivalent to that obtained from (\ref{W2}) was studied in a non-supersymmetric framework. 
In a special case, the inflaton dynamics is equivalent to that in the spontaneous 
symmetry breaking model first considered in \cite{Linde:1983fq} (see also Ref.~\cite{Kallosh:2007wm}).
Recently, the  model was revisited and its global supersymmetric extension was proposed in Ref.~\cite{Croon:2013ana},
where the predicted spectral index and the tensor-to-scalar ratio can be
consistent with the Planck data, although
the possible inflationary trajectory is limited to the case of Ref.~\cite{Linde:1983fq} as the inflaton 
is a complex scalar field. The extension to SUGRA is possible in the no-scale 
supergravity~\cite{Cremmer:1983bf,Ellis:1983sf,Murayama:1993xu}. See also a very recent paper~\cite{Ellis:2013xoa}
along those lines.

In this paper we present a general polynomial chaotic inflation model in SUGRA as an extension of our previous work~\cite{Nakayama:2013jka},
and show that the Planck result can be consistent with a large class of polynomial chaotic inflation models.
Most importantly, the predicted tensor-to-scalar ratio is large enough to be probed by the on-going and 
future experiments dedicated to detect the CMB B-mode polarization signal. 
Conversely, non-detection of B-modes at future experiments will exclude a large portion of
the polynomial chaotic inflation model.

The rest of this paper is organized as follows.
In Sec.~\ref{sec:pol} we introduce the setup of polynomial chaotic inflation model in SUGRA.
We will discuss the reheating process in Sec.~\ref{sec:reh}. The last section is devoted for
discussion and conclusions.

%%%%%%%%%%%%%%%%%%%%%%%%%%%%%%%%%%%%%
\section{General polynomial chaotic inflation}   \label{sec:pol}
%%%%%%%%%%%%%%%%%%%%%%%%%%%%%%%%%%%%%

In this section we present a general polynomial chaotic inflation in SUGRA,
and study the predicted $(n_s, r)$ in a few examples.

We consider the following K\"ahler potential,
\begin{equation}
	K = c_1(\phi+\phi^\dagger) + \frac{1}{2}(\phi+\phi^\dagger)^2+ |X|^2 - c_X|X|^4 \cdots,
	\label{K}
\end{equation}
which respects the shift symmetry (\ref{shift}).  The dots represent higher order terms that are not
relevant for the present purpose. We assume that  the coefficients of interactions in the K\"ahler potential 
are at most of order unity.
The shift symmetry ensures the flatness of the scalar potential beyond the Planck scale 
along the imaginary component of $\phi$,  $\varphi \equiv \sqrt{2}\,{\rm Im}\phi$, and 
hence it allows us to identify $\varphi$ with the inflaton.
The superpotential is assumed to be of the general
form~\cite{Kawasaki:2000yn,Kawasaki:2000ws,Kallosh:2010ug,Kallosh:2010xz},
\begin{equation}
	W = X \left(\sum_{n=0} d_n \phi^n \right)
	 + W_0,
	\label{W_gen}
\end{equation}
where $\{d_n\}$ are constants  and $|W_0| \simeq m_{3/2}$ with $m_{3/2}$ being the gravitino mass.
Here we assign an R-charge as $R(X)=2$ and $R(\phi)=0$.
The superpotential that involves $\phi$ explicitly breaks the shift symmetry and generates 
the potential for $\varphi$ unless there is a conspiracy among the coefficients.\footnote{
\label{ft}
The shift symmetry is not violated if $\sum_{n=0} d_n \phi^n \propto e^{c \phi}$ with $c$ being a real
parameter, because it leads to the scalar potential $V \propto e^{c(\phi+\phi^\dagger)}$ in the R-symmetric limit.
}
The scalar potential is given by
\begin{equation}
	V \;=\; e^K\left[ K^{i\bar j} (D_iW)(D_{\bar j}\bar W) -3|W|^2 \right],
	\label{Vsugra}
\end{equation}
where $D_iW = \partial_i W + (\partial_i K)W $.

Now let us make a couple of simplifying assumptions. First we assume that the gravitino mass is much
smaller than the Hubble parameter during inflation. Then the effect of $W_0$ on the inflaton dynamics
is negligible, and so, we drop $W_0$ in the rest of this section. Since there is an approximate U(1)$_R$ symmetry
during inflation, the potential of $X$ has an  extremum at the origin. In fact, 
$X$ dynamically settles down at $X \simeq 0$ during inflation if $c_X \gtrsim \mathcal O(1)$,
which greatly simplifies the scalar potential. In fact,  
the introduction of $X$ was essential for successful inflation, since otherwise the the potential would be
unbounded from below at large $\varphi$~\cite{Kawasaki:2000yn}.
Secondly we assume that the scalar potential has a (SUSY) minimum at the origin of $\phi$.
There are two ways to accomplish this. First, the potential vanishes at the origin
if the $d_0$-term is forbidden by an additional symmetry.  We shall see that there is an interesting class of
models where we can impose a $Z_2$ symmetry on $X$ and $\phi$. 
Alternatively, we can use the freedom to shift the origin of $\phi$. In general, this is possible if
$|d_0|$ is  smaller than or comparable to the typical scale of $\{d_i\}$ with $i\geq 1$.
Otherwise, we cannot take $d_0=0$ by the shift of $\phi$
since ${\rm Re}[\phi]$ is not allowed to take super-Planckian values.\footnote{If the phase of $d_0$ happens to be
correlated with the $d_i$-terms, it will be possible to shift the origin along the imaginary component
of $\phi$ even for $|d_0| \sim {\cal O}(1)$.}
%
%If the $d_0$-term is not
%forbidden by any symmetry, we naively expect $|d_0| \sim {\cal O}(1)$.  On the other hand,
%$\{d_i\}$ with $i\geq 1$ are considered to be much smaller than unity as they represent
%the explicit breaking of the shift symmetry. Then, the shift is possible only if ${\rm Re}[d_0]$
%is  smaller than or comparable to the typical scale of $\{d_i\}$. This is because 
%
%
%\footnote{
%	{\bf  
%	Note that the flatness of $\varphi$ is ensured up to $\varphi \lesssim \mathcal O(m^{-1})$ with $m$ being the inflaton mass,
%	because the K\"ahler potential may have the shift symmetry breaking term of the order $K \sim m^2|\phi|^2$.
%	Therefore, the shift of $\phi$ to eliminate the $d_0$ term is justified as long as $|d_0| \lesssim \mathcal O(1)$.
%	}
%}
%Note, however,  that if $\phi$ has a non-trivial symmetry other than the shift symmetry, it becomes
%obscured after the shift, unless the $d_0$-term is forbidden by the symmetry. 
In this basis, $c_1$ is generically at most of order unity,  and there is a priori no reason to expect $c_1$ to be 
much smaller than unity. 
Nevertheless we will set $c_1 = 0$ for simplicity in the following analysis, because it does not affect
the inflaton dynamics (see the footnote~\ref{ft}).
In the next section we will return to this issue and
we shall see that $c_1 = 0$ can be realized in some cases. 
Other phenomenological implications of nonzero $c_1$ will also be discussed there.

Under the above assumptions we have made, the inflaton potential takes the simple form,
\beq
V(\varphi)\;\simeq\;\left| \sum_{n=1} d_n \lrfp{i \varphi}{\sqrt{2}}{n} \right|^2,
\label{Vinf}
\eeq
where we have used the fact that the real component of $\phi$ is stabilized near the origin
during inflation. 
In the chaotic inflation, the inflaton rolls down toward the potential minimum from large
field values, and so, different terms in the superpotential will give a 
dominant contribution to the inflaton potential as the inflaton moves. 
If only a single term gives the dominant contribution during the last $50-60$ e-foldings,
the inflation model is reduced to that based on a monomial potential. 
Instead, we would like to consider a case where during the last $50-60$ e-foldings 
the inflaton passes through the region where the inflaton potential receives comparable contributions from
(at least) two terms in the superpotential. As we shall see shortly, this will significantly change
the predicted values of $n_s$ and $r$. This is because the scalar potential has a  plateau where
these two contributions compete and cancel each other. 
In the following we first study such a general inflation model,
and then examine a couple of specific examples to see if the predicted $(n_s, r)$ can 
lie within the $1\sigma$ region allowed by the Planck.

%%%%%%%%%%%%%%%%%%%%%%%%%%%%%%%%%%%%%
\subsection{$(p, q)$-chaotic inflation}
%%%%%%%%%%%%%%%%%%%%%%%%%%%%%%%%%%%%%

%%%%%%%%%%%%%%%%
\begin{figure}
\begin{center}
\includegraphics[scale=1.2]{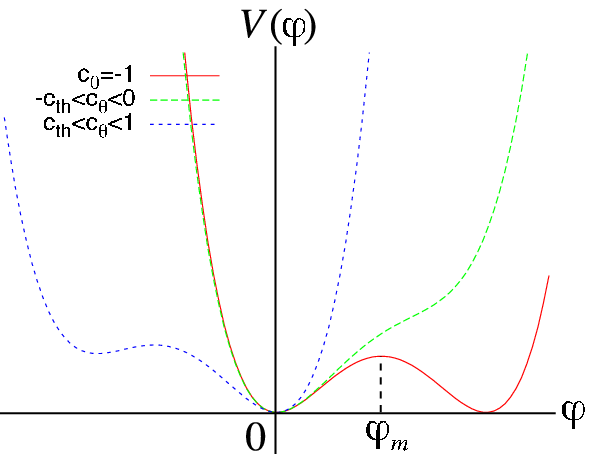}~~~
\includegraphics[scale=1.2]{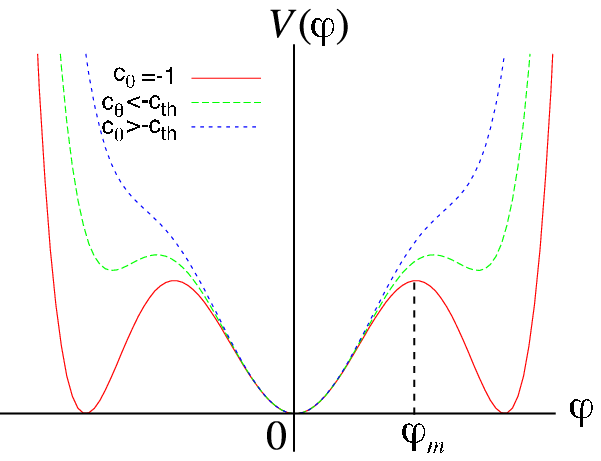}
\caption{ 
	Schematic picture of the inflaton potential for the general $(p,q)$-chaotic inflation. 
	Here $c_\theta \equiv \cos(\theta+(q-p)\pi/2)$ and $c_{\rm th} \equiv 2\sqrt{pq}/(p+q)$.
	(Left) $q-p=$odd. (Right) $q-p=$even.
}
\label{fig:potpq}
\end{center}
\end{figure}
%%%%%%%%%%%%%%%%

Now we consider a $(p,q)$-chaotic inflation in which the inflaton potential  receives
contributions mainly from the following two terms in the superpotential during the last $50-60$ e-foldings;
\begin{equation}
	W = X (d_p \phi^p + d_q \phi^q), 
	\label{Wpq}
\end{equation}
where $p$ and $q$ are integers satisfying $1 \leq p < q$. We take $d_p$ real and positive without loss of generality.
Using   (\ref{Vinf}) and (\ref{Wpq}), we find
\begin{equation}
	V =d_p^2 \lrfp{\varphi}{\sqrt{2}}{2p} %\frac{}{2^p}\varphi^{2p}
	\left\{
		1+ 2 % \cos\left(\theta+\frac{(q-p)\pi}{2}\right)  \xi %\left|\frac{d_q}{d_p}\right|
		\cos\Theta   \,\xi %\left|\frac{d_q}{d_p}\right|
		\lrfp{\varphi}{\sqrt{2}}{q-p}
		%\frac{d}{2^{(q-p-2)/2}} \varphi^{q-p} 
		+ \xi^2 %\left|\frac{d_q}{d_p}\right|^2
		 \lrfp{\varphi}{\sqrt{2}}{2(q-p)}
	\right\},
\end{equation}
where we have defined $\xi \equiv |d_q/d_p|$,
$\theta \equiv {\rm arg}(d_q) $ and $\Theta \equiv \theta + (q-p)\pi/2$.

%Hereafter we consider the case of $p > 0$.
In order to see how the scalar potential looks like, let us first consider the case of $\cos \Theta=-1$,
which is shown as the solid (red) line in Fig.~\ref{fig:potpq}. The shape of the potential depends on 
$q-p$ being odd or even. If $q-p$ is odd, the potential has three extrema at $\varphi = 0, \varphi_m$ 
and $(q/p)^{1/(q-p)} \varphi_m$, where
\begin{equation}
	\varphi_m = \sqrt{2}\left( \frac{p}{q \xi} \right)^{1/(q-p)},
\end{equation}
On the other hand, if $q-p$ is even, the potential is symmetric under $\varphi \to -\varphi$
and there are five extrema at $\varphi = 0$, $\pm\varphi_m$ and $\pm(q/p)^{1/(q-p)} \varphi_m$.
Schematic pictures of this potential are given in Fig.~\ref{fig:potpq} for $q-p =$\,odd (left) and $q-p =$\,even (right).

For a general value of $\Theta$,  the potential and its first derivative with respect to $\varphi$,   can be rewritten as
\begin{equation}
	V =d_p^2 \lrfp{\varphi}{\sqrt{2}}{2p}\left\{
		1+ 2 \cos \Theta \,\frac{p}{q}\left(\frac{\varphi}{\varphi_m}\right)^{q-p} 
		+\left(\frac{p}{q}\right)^{2}  \left(\frac{\varphi}{\varphi_m}\right)^{2(q-p)} 
	\right\},
\end{equation}
and
\begin{equation}
	V' = \sqrt{2} d_p^2\, p\lrfp{\varphi}{\sqrt{2}}{2p-1}%\frac{p d_p^2}{2^{p-1}}\varphi^{2p-1}
	 	\left\{
		1+ \cos\Theta\, \frac{p+q}{q} \left(\frac{\varphi}{\varphi_m}\right)^{q-p} 
		+\frac{p}{q}  \left(\frac{\varphi}{\varphi_m}\right)^{2(q-p)} 
	\right],
\end{equation}
where the prime denotes the the derivative with respect to the inflaton. 

It is also instructive to write down the condition for the extrema other than $\varphi \neq 0$ to disappear.
If the following condition is satisfied,
\begin{equation}
\begin{split}
	\left| \cos \Theta \right| < \frac{2\sqrt{pq}}{p+q}& ~~~{\rm for}~~~q-p={\rm odd},\\
	\cos \Theta > -\frac{2\sqrt{pq}}{p+q}& ~~~{\rm for}~~~q-p={\rm even},
	\label{nomin}
\end{split}
\end{equation}
$V'=0$ has a solution only at $\varphi=0$ and hence there is no local minimum for the inflaton potential,
and inflation naturally takes place with chaotic initial conditions as in the original chaotic inflation models.
When the above condition is marginally satisfied, the inflaton potential has a plateau around $\varphi = \varphi_m$.

One can see from Fig.~\ref{fig:potpq}  that  the potential is approximated by a simple monomial potential, $V \propto \varphi^{2p}$, for $\varphi \ll \varphi_m$,
while the potential becomes significantly modified at $\varphi \gtrsim \varphi_m$. In particular, 
 a plateau-like feature appears around $\varphi \sim \varphi_m$ depending on the value of $\theta$.
Thus,  the inflaton dynamics and the predicted spectral index and tensor-to-scalar ratio
are significantly modified. To see this, let us remember
that the spectral index as well as the scalar-to-tensor ratio can be expressed in terms of the
slow-roll parameters as~\cite{Liddle:2000cg}
\bea
\label{ns}
n_s &=& 1 + 2 \eta - 6\epsilon \\
\label{r}
r&=& 16 \epsilon,
\eea
where the slow-roll parameters $\epsilon$ and $\eta$ are defined by
\bea
\epsilon &\equiv& \frac{1}{2} \lrfp{V'}{V}{2},\\
\eta &\equiv & \frac{V''}{V}.
\eea
For illustration we consider the simple quadratic chaotic inflation. We can see from Fig.~\ref{fig:p1q2} that
the predicted $r$ is on the edge of the $2 \sigma$  region, while $n_s$ is  close
to the best-fit value. From the above expressions, we need to reduce $\epsilon$ to suppress $r$.
However, this will make the spectral index larger. So, in order to effectively reduce only $r$, we
need to make both $\epsilon$ and $\eta$ smaller at $\varphi \simeq {\cal O}(10)$ where the CMB scales
exited the horizon. That is to say, the potential should be flatter and
its curvature should be small and even negative. Such deformation of the potential at large field values 
is possible in the polynomial chaotic inflation, if the condition (\ref{nomin}) is marginally satisfied.

Next we will see some examples and show that they indeed predict $n_s$ and $r$ within 
the 1$\sigma$ region favored by the Planck satellite.

%%%%%%%%%%%%%%%%%%%%%%%%%%%%%%%%%%%%%
\subsection{Examples}  \label{sec:ex}
%%%%%%%%%%%%%%%%%%%%%%%%%%%%%%%%%%%%%

Now let us study some phenomenological examples of the polynomial chaotic inflation
to see if they predict $n_s$ and $r$ within the $1\sigma$ allowed region.

%%%%%%%%%%%%%%%%%%%%%%%%%%%%%%%%%%%%%
\subsubsection{$(1, 2)$-chaotic inflation}
%%%%%%%%%%%%%%%%%%%%%%%%%%%%%%%%%%%%%

%%%%%%%%%%%%%%%%
\begin{figure}
\begin{center}
\includegraphics[scale=1.2]{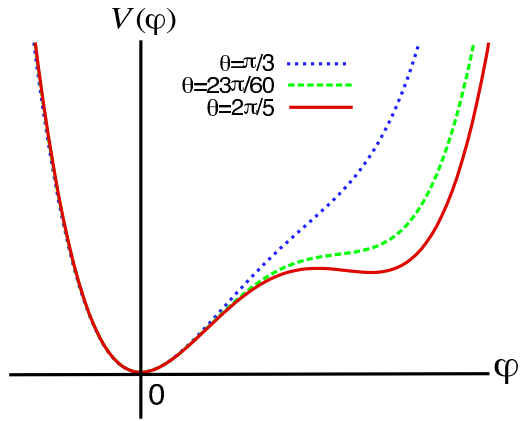}
\caption{ 
	Schematic picture for the scalar potential for $(p,q)=(1,2)$.
}
\label{fig:pot12}
\end{center}
\end{figure}
%%%%%%%%%%%%%%%%

%%%%%%%%%%%%%%%%
\begin{figure}
\begin{center}
\includegraphics[scale=1.2]{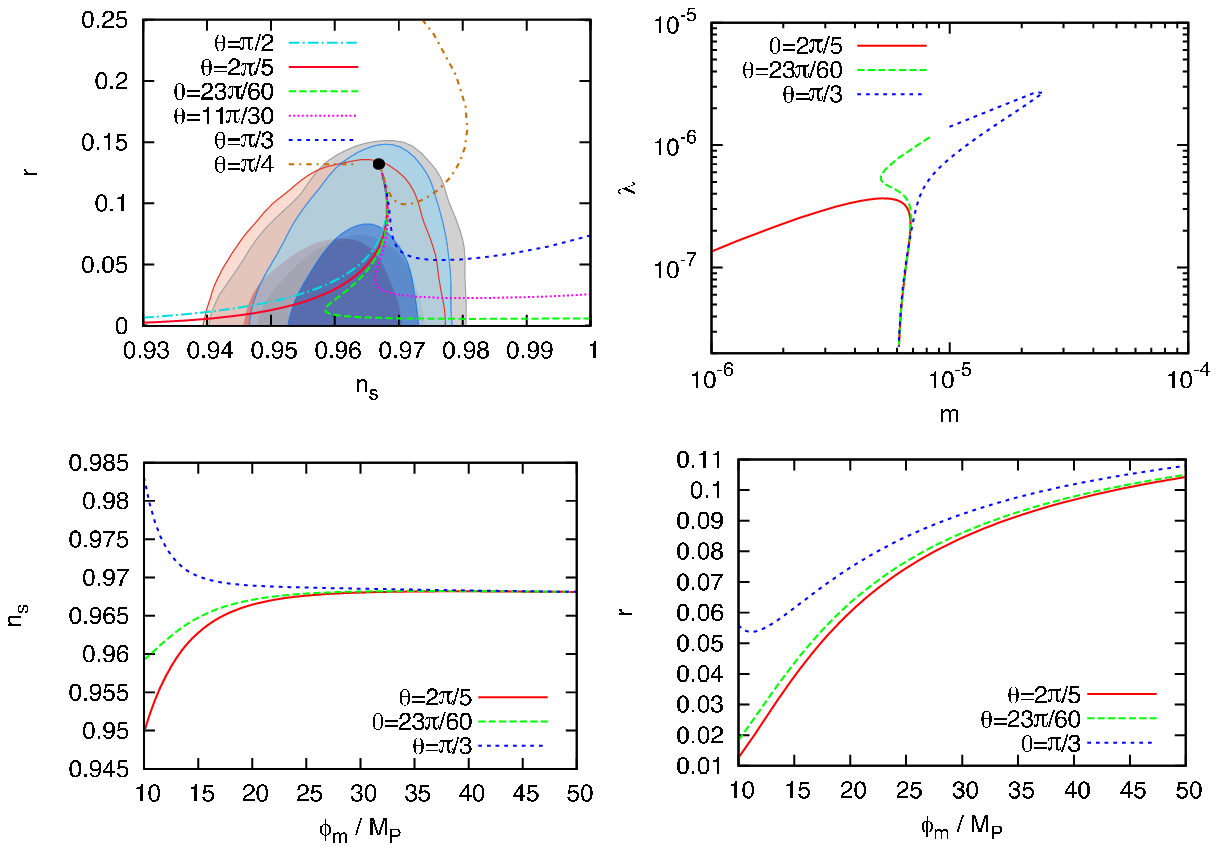}
\caption{ 
	Predictions of the polynomial chaotic inflation with $(p,q)=(1,2)$ for various values of $\theta$.
	Upper left : Predictions for $(n_s,r)$ combined with the Planck constraint~\cite{Ade:2013rta}.
	Upper right : Values of $m$ and $\lambda$ (in Planck units) for reproducing the Planck normalization of the density perturbation.
	Lower left : $n_s$ as a function of $\varphi_m$.
	Lower right : Same as lower left, but for $r$.
}
\label{fig:p1q2}
\end{center}
\end{figure}
%%%%%%%%%%%%%%%%

The simplest case is $(p,q) = (1,2)$, which was studied in Ref.~\cite{Nakayama:2013jka}.
Writing $d_p = m$ and $|d_q|=\lambda$, we have the superpotential
\begin{equation}
	W = X(m\phi + \lambda e^{i\theta} \phi^2),
\end{equation}
and the scalar potential becomes
\begin{equation}
	V = \frac{1}{2}m^2\varphi^2 \left( 1-\frac{\sqrt{2}\lambda\sin\theta}{m}\varphi 
	+ \frac{\lambda^2}{2m^2}\varphi^2 \right).
\end{equation}
The potential has a plateau when the following condition is marginally satisfied,
\begin{equation}
	|\sin \theta| < \frac{2\sqrt{2}}{3}.  \label{theta12}
\end{equation}
Schematic pictures for the scalar potential are shown in Fig.~\ref{fig:pot12}.
As seen from the figure, the potential has a flat plateau for $\theta \sim 23\pi/60$, which marginally satisfies (\ref{theta12}).
Note that in this case the symmetry arguments do not forbid the $c_1$ term in the K\"ahler potential (\ref{K}),
and non-zero value of $c_1$ will induce the inflaton decay through supergravity effects as we shall see later.
%and $d_0$ term in the superpotential (\ref{W_gen}), both of which are absorbed into the redefinition of the field $\phi \to \phi' = \phi + {\rm const.}$ and parameters $m$ and $\lambda$.

We have numerically solved the inflaton dynamics and calculated $n_s$ and $r$ in this model,
which can be expressed in terms of the slow-roll parameters as (\ref{ns}) and (\ref{r}).
They are evaluated at $\varphi = \varphi (N_e)$ where $\varphi (N_e)$ satisfies
\begin{equation}
	N_e = \int_{\varphi_{\rm end}}^{\varphi(N_e)} \frac{V}{V'}d\varphi,
\end{equation}
with $N_e$ e-folds before the end of inflation, and $\varphi_{\rm end}$ is defined at the field value where
${\rm max}[\epsilon, |\eta|] = 1$.
For numerical calculation, we have taken $N_e=60$.
Upper left panel of Fig.~\ref{fig:p1q2} shows predictions of the polynomial chaotic inflation on $(n_s,r)$ plane for various values of $\theta$ overlapped with the 1$\sigma$ and 2$\sigma$ region from the Planck results~\cite{Ade:2013rta}.
The filled black circle shows the prediction for the chaotic inflation with quadratic potential.
We can see that the predicted values of $(n_s, r)$ can lie within the $1\sigma$ allowed region. 
Unless $\theta$ is finely tuned around $\theta = 23\pi/60$, the predicted
tensor-to-scalar ratio is large enough to be probed by the on-going and future B-mode experiments.
Lower panels of Fig.~\ref{fig:p1q2} show $n_s$ and $r$ as a function of $\varphi_m$ for various values of $\theta$.
For $\varphi_m \gg 10$, the prediction is reduced to that for the chaotic inflation with a quadratic potential, as  expected.
We have also checked that the correct magnitude of the density perturbation observed by the Planck satellite,
$\sqrt{\mathcal P_\zeta} = \sqrt{V/(24\pi^2\epsilon)} \simeq 4.69\times 10^{-5}$~\cite{Ade:2013rta},
is reproduced for $m \sim 10^{-5}$ and $\lambda \sim 10^{-7}$, as shown in the upper right panel of Fig.~\ref{fig:p1q2}.

%%%%%%%%%%%%%%%%%%%%%%%%%%%%%%%%%%%%%
\subsubsection{$(1, 3)$-chaotic inflation}
%%%%%%%%%%%%%%%%%%%%%%%%%%%%%%%%%%%%%

%%%%%%%%%%%%%%%%
\begin{figure}
\begin{center}
\includegraphics[scale=1.2]{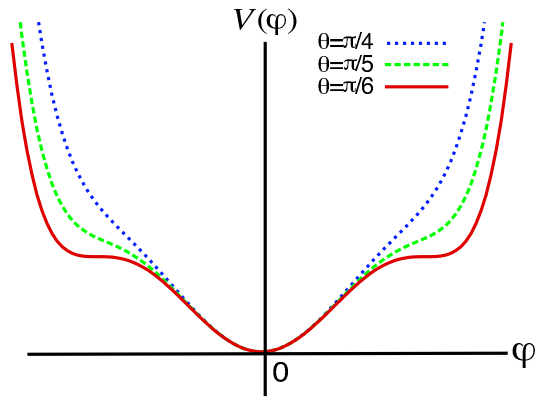}
\caption{ 
	Schematic picture for the scalar potential for $(p,q)=(1,3)$.
}
\label{fig:pot13}
\end{center}
\end{figure}
%%%%%%%%%%%%%%%%

%%%%%%%%%%%%%%%%
\begin{figure}
\begin{center}
\includegraphics[scale=1.2]{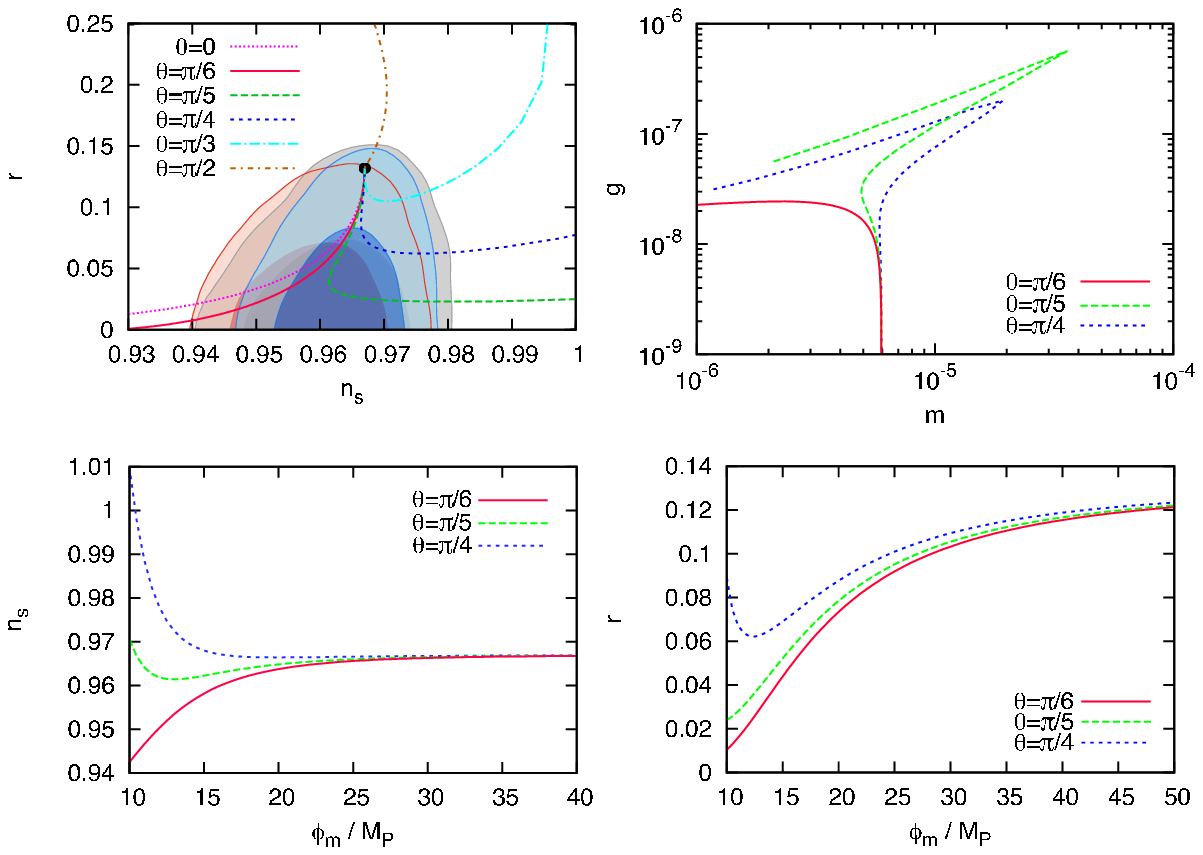}
\caption{ 
	Predictions of the polynomial chaotic inflation with $(p,q)=(1,3)$ for various values of $\theta$.
	Upper left : Predictions for $(n_s,r)$. 
	Upper right : Values of $m$ and $g$ (in Planck units) for reproducing the Planck normalization of the density perturbation.
	Lower left : $n_s$ as a function of $\varphi_m$.
	Lower right : Same as lower left, but for $r$.
}
\label{fig:p1q3}
\end{center}
\end{figure}
%%%%%%%%%%%%%%%%

Another simple interesting possibility is the case of $(p,q)=(1,3)$.
Writing $d_p=m$ and $|d_q|\equiv g$, the superpotential is given by
\begin{equation}
	W = X(m\phi + g e^{i\theta}\phi^3).
	\label{W13}
\end{equation}
%%
%This form of the superpotential is consistent with the $Z_2$-symmetry under 
%which $X\to -X$ and $\phi \to -\phi$.
The scalar potential becomes
\begin{equation}
	V = \frac{1}{2}m^2\varphi^2 \left( 1-\frac{g\cos\theta}{m}\varphi^2 + \frac{g^2}{4m^2}\varphi^4 \right).
\end{equation}
The condition (\ref{nomin}) reads
\begin{equation}
	\cos \theta < \frac{\sqrt{3}}{2}.
	\label{13}
\end{equation}

Fig.~\ref{fig:p1q3} shows the prediction of $(n_s,r)$ for various values of $\theta$
combined with the 1$\sigma$ and 2$\sigma$ region from the Planck results~\cite{Ade:2013rta}.
The filled black circle shows the prediction for the chaotic inflation with quadratic potential.
It is clearly seen that it lies within the 1$\sigma$ favored region for $\theta$ around $\pi/5$,
which marginally satisfies the condition (\ref{13}).

From a phenomenological point of view,   the case of $(p,q)=(1,3)$ is particularly attractive, 
since it is consistent with the $Z_2$ symmetry  under 
which $X\to -X$ and $\phi \to -\phi$. The $Z_2$ symmetry forbids the $d_0$ term and
there is no need to shift the origin of $\phi$ to ensure the superpotential (\ref{W13}).
Furthermore, the $c_1$-term in the K\"ahler potential is also forbidden by the symmetry.
As we shall see in the next section,  we do not suffer from the gravitino overproduction from the inflaton decay.
The reheating temperature can be as high as $T_{\rm R} \sim 10^9$\,GeV by the inflaton decay into 
right-handed neutrinos or  Higgs fields.

%%%%%%%%%%%%%%%%%%%%%%%%%%%%%%%%%%%%%
\subsubsection{$(2, 4)$-chaotic inflation}
%%%%%%%%%%%%%%%%%%%%%%%%%%%%%%%%%%%%%

Lastly we consider the case of $(p,q)=(2,4)$ for illustration.
Writing $d_p=\lambda$ and $|d_q|\equiv g$, the superpotential is of the form
\begin{equation}
	W = X(\lambda\phi^2 + g e^{i\theta}\phi^4).
\end{equation}
The scalar potential is given by
\begin{equation}
	V = \frac{1}{4}\lambda^2\varphi^4 \left( 1-\frac{g\cos\theta}{\lambda}\varphi^2 + \frac{g^2}{4\lambda^2}\varphi^4 \right).
	\label{Vp2q4}
\end{equation}
The condition (\ref{nomin}) becomes
\begin{equation}
	\cos \theta < \frac{2\sqrt{2}}{3}.
\end{equation}

\begin{figure}
\begin{center}
\includegraphics[scale=1.2]{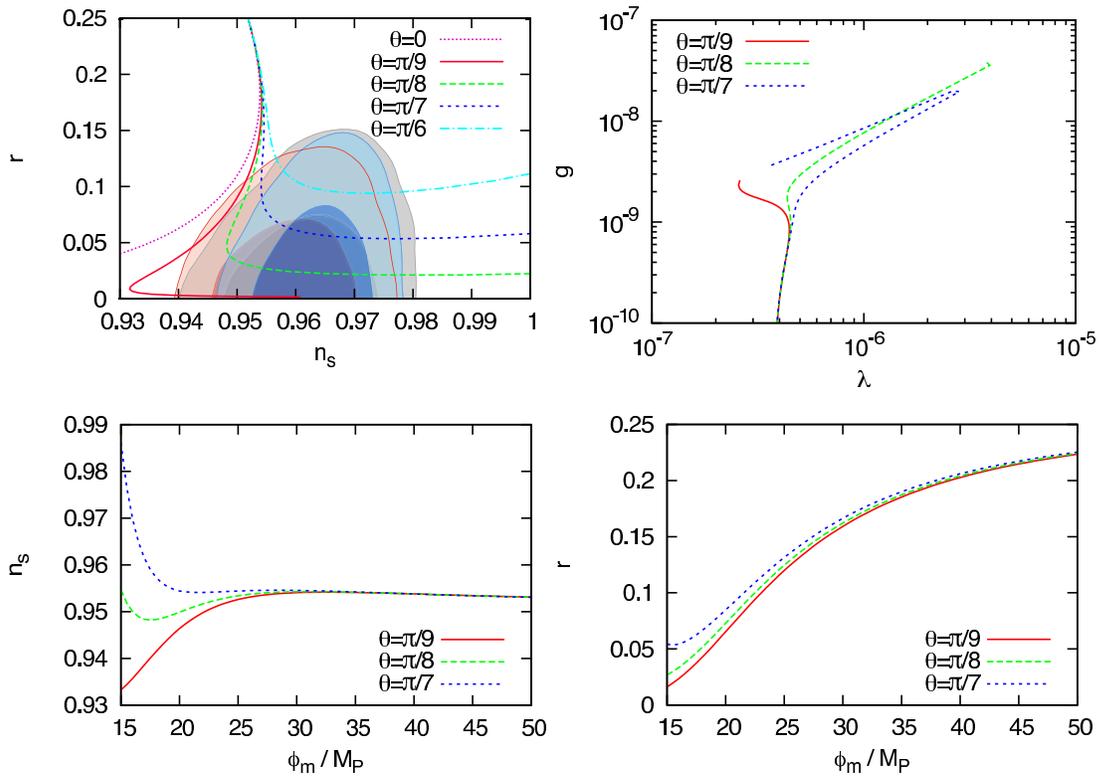}
\caption{ 
	Predictions of the polynomial chaotic inflation with $(p,q)=(2,4)$ for various values of $\theta$.
	Upper left : Predictions for $(n_s,r)$. 
	Upper right : Values of $\lambda$ and $g$ (in Planck units) for reproducing the Planck normalization of the density perturbation.
	Lower left : $n_s$ as a function of $\varphi_m$.
	Lower right : Same as lower left, but for $r$.
}
\label{fig:p2q4}
\end{center}
\end{figure}

The lowest order term of the inflaton potential is proportional to $\varphi^4$, which would be
 strongly disfavored from the Planck result if it gave the dominant contribution to the inflaton potential
 during the relevant time period. Now, as the potential is modified by the higher order terms,
the predicted $(n_s, r)$ can be consistent with the Planck result 
as shown in Fig.~\ref{fig:p2q4}.
This model is also consistent with the $Z_2$ symmetry under which $\phi\to-\phi$ and $X\to X$,
hence the $c_1$ term in the K\"ahler potential (\ref{K}) can be forbidden.
Note however that the $d_0$ term in the superpotential (\ref{W_gen}), $W \supset X d_0$, is allowed,
and if we shift the origin of $\phi$ to cancel the $d_0$-term, the structure of the potential would
be modified; for instance terms like $X\phi$ and $X\phi^3$  would be induced after the shift. 
 If $|d_0| \ll \lambda^2/g$ is satisfied from the beginning, 
the effect of the $d_0$-term is so small that the inflaton dynamics is not affected.
We have here assumed that $|d_0|$ is sufficiently suppressed in our analysis.

%%%%%%%%%%%%%%%%
\begin{figure}
\begin{center}
\includegraphics[scale=1.2]{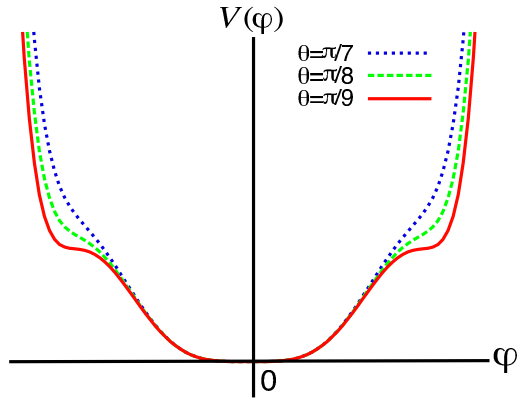}
\caption{ 
	Schematic picture for the scalar potential for $(p,q)=(2,4)$.
}
\label{fig:pot24}
\end{center}
\end{figure}
%%%%%%%%%%%%%%%%

%%%%%%%%%%%%%%%%
%%%%%%%%%%%%%%%%

%%%%%%%%%%%%%%%%%%%%%%%%%%%%%%%%%%%%%
\section{Reheating}  \label{sec:reh}
%%%%%%%%%%%%%%%%%%%%%%%%%%%%%%%%%%%%%

In this section we study the inflaton decay. So far we have neglected the linear term of the
inflaton in the K\"ahler potential as it does not affect the inflaton dynamics. However, it has
an important effect on the inflaton decay process. This can be easily seen by the K\"ahler
transformation: $K \to K - c_1 (\phi+\phi^\dag)$ and $W \to e^{c_1 \phi} W$. Then, the inflaton
tadpole disappears in the K\"ahler potential, but the inflaton is now coupled to all the sectors
appearing in the superpotential. Such couplings will induce the inflaton decay into the 
visible sector through the top Yukawa coupling~\cite{Endo:2006qk}. 

If we use the freedom to shift the origin of $\phi$ in order to absorb
the $d_0$ term in the superpotential, it generically induces a non-zero $c_1$-term.
On the other hand, the $d_0$ term can be forbidden by some symmetry under which
 both $X$ and $\phi$ are charged. This implies that $c_1$ term can also be forbidden by the symmetry.
As long as we require the $(\phi+\phi^\dag)^2$ term in the K\"ahler potential to be
consistent with such symmetry, the only possible symmetry is the $Z_2$ symmetry
under which $X \to -X$ and $\phi \to - \phi$.\footnote{
If the kinetic term of the inflaton comes from higher order terms at large field values,
it is possible to impose a $Z_n$ symmetry consistently. This is the case of the so called
running kinetic inflation~\cite{Takahashi:2010ky,Nakayama:2010kt}.
} The $Z_2$ symmetry forbids all the terms
with $n$ being even in (\ref{W_gen}). If we focus on the case where the first leading two
terms are relevant, we obtain the $(1,3)$-chaotic inflation. In this sense, the $(1,3)$-model
is special. We discuss below the inflaton decay process in cases with and without such
$Z_2$ symmetry. 

%%%%%%%%%%%%%%%%%%%%%%%%%%%%%%%%%%%%%
\subsection{Case with $Z_2$ symmetry}
%%%%%%%%%%%%%%%%%%%%%%%%%%%%%%%%%%%%%
Now let us consider the reheating in the $Z_2$-symmetric model, e.g., $(p,q)=(1,3)$.
In this case the $c_1$-term vanishes, and therefore we need to introduce couplings
of the inflaton with the visible sector for successful reheating. As we shall see, however,
the couplings are bounded above, either because too large $c_1$ term is induced by the couplings,
or because the inflaton trajectory is destabilized. This leads to an upper bound
on the reheating temperature, which however turns out to be  sufficiently high for thermal or non-thermal
leptogenesis~\cite{Fukugita:1986hr,Lazarides:1991wu,Asaka:1999yd} to work. 
Another phenomenologically attractive feature of the $Z_2$ symmetric model is that the gravitino overproduction 
from inflaton decay can be suppressed. This is because any couplings of the inflaton with the SUSY breaking
sector are suppressed by the $Z_2$ symmetry, as long as its breaking is sufficiently small. 
%as long as all the particles in the SUSY breaking sector are neutral under the $Z_2$. 

In the following we assume $p=1$
so that the inflaton acquires a large SUSY mass, $m$,  at the potential minimum. 
That is to say, the lowest order term in the superpotential is $W = mX \phi$. 
As we consider a perturbative decay of the inflaton, higher order terms are irrelevant for the reheating. 

%%%%%%%%%%%%%%%%%%%%%%%%%%%%%%%%%%%%%
\subsubsection{Decay into the right-handed neutrino}
%%%%%%%%%%%%%%%%%%%%%%%%%%%%%%%%%%%%%

Let us consider the inflaton coupling to the right-handed neutrinos $N_i$ $(i=1,2,3)$ :
\begin{equation}
	W = mX\phi + Y_{ij} m \phi N_i N_j + \frac{1}{2}\kappa_i \Pi  N_i N_i,
	\label{phiNN}
\end{equation}
where  $Y_{ij}$ and $\kappa_i$ denote  coupling constants, and a sum over the generation index 
$i, j=1,2,3$ is understood.
The VEV of $\Pi $ generates  a Majorana mass for the right-handed neutrinos,
$M_i \equiv \kappa_i \la \Pi \ra$. 
The second term explicitly breaks the shift symmetry, and therefore 
we have inserted an order parameter for the breaking, $m$. 

In order to allow the above interactions, we extend the $Z_2$ symmetry to a $Z_4$ symmetry.  See the  
%The model can be consistent with the $Z_4\times$U(1)$_R$ symmetry with 
charge assignments in Table~\ref{tab:charge}~\cite{Kawasaki:2000ws}.
This $Z_4$ symmetry is broken down to its $Z_2$ subgroup by the VEV of $\Pi $.
As a result, a non-zero inflaton VEV of order $O(\la \Pi \ra/m)$ is induced, as the $d_0$-term is 
generically of order $\la \Pi \ra$. 
We assume $\la \Pi \ra \ll m$ so that the inflaton VEV is much smaller than unity, in which case the
effect on the reheating is negligible. Then the inflaton  can decay into the three right-handed neutrinos.

We first consider the inflaton decay into the lightest  one, $N_1$.
The decay into heavier ones can be treated in the same way.
The superpotential (\ref{phiNN}) leads to the following terms relevant for the reheating,
\begin{equation}
	\mathcal L = -(Ym\phi N_1 N_1 + {\rm h.c.}) - (Y m^2 X^\dagger \tilde N_1 \tilde N_1 +{\rm h.c.}),
	\label{intN}
\end{equation}
where $Y \equiv Y_{11}$, and $N_1$ and $\tilde{N}_1$ represent the right-handed neutrino and sneutrino,
respectively. 

After inflation, $\phi$ and $X$ are maximally mixed with each other due to the constant term $W_0$ in the superpotential
(see Eq.~(\ref{W_gen})) and form the mass eigenstates as $\Phi_{\pm} = (\phi \mp X^\dagger)/\sqrt{2}$ with mass eigenvalues $m^2_{\pm} = m(m\pm m_{3/2})$~\cite{Kawasaki:2006gs}. As long as the inflaton decay rate
is much smaller than $m_{3/2}$, both mass eigenstates are equally populated. In addition, their couplings with the
right-handed neutrinos are almost equal in magnitude, and we do not have to distinguish them for the present purpose. 

The inflaton decay rate through the interactions (\ref{intN}) is then given by
\begin{equation}
	\Gamma( \Phi \to N_1N_1) \simeq \Gamma(\Phi\to\tilde N_1\tilde N_1) \simeq \frac{|Y|^2}{16\pi}m^3.
\end{equation}
Assuming the same decay rate for $\phi \to N_2 N_2$ and $\phi \to N_3 N_3$, 
we obtain the reheating temperature as\footnote{
	Typically the decay rate of the right-handed (s)neutrino is much larger than the inflaton decay rate,
	hence the produced right-handed (s)neutrinos decay immediately after the inflaton decay.
}
\begin{equation}
	T_{\rm R} \simeq 3\times 10^9\,{\rm GeV}~|Y|
	\left( \frac{m}{10^{13}\,{\rm GeV}} \right)^{3/2}.
	\label{TR_N}
\end{equation}
Therefore it can be as high as $10^9$--$10^{10}$\,GeV for $|Y| \sim \mathcal O(1)$\footnote{
Note that $|Y|$ cannot be larger than order unity, as it would destabilize the inflaton trajectory.
See the discussion in Sec.~\ref{sec:higgs}.
},
and thermal and/or non-thermal leptogenesis works successfully in this case.

%%%%%%%%%%%%%%
\begin{table}[t]
\begin{center}
  \begin{tabular}{c|cccccccc}
    \hline
    \hline
                    &$\phi$ & $X$ & $\Pi $ & $N$ & $H_u$ & $H_d$ & ${\bf 10}$ & $\bar {\bf 5}$\\
    \hline
    $Z_4$       & $2$ & $2$ & $2$ &$1$ & $2$ & $2$ & $1$ & $1$\\
    \hline
    U(1)$_R$ & $0$ & $2$ & $0$ &$1$ & $0$ & $0$ & $1$ & $1$\\
    \hline
  \end{tabular}
\label{tab:charge}
\end{center}
\caption{ $Z_4\times$U(1)$_R$ charge assignments. Here ${\bf 10} =(Q, \bar U, \bar E)$ and $\bar {\bf 5} = (\bar D, L)$
are MSSM chiral matters.}
\label{default}
\end{table}
%%%%%%%%%%%%%%

It is also possible to introduce the following coupling in the K\"ahler potential consistent with the $Z_4\times$U(1)$_R$:
\begin{equation}
	K = k X^\dagger N_i N_i + {\rm h.c.}
\end{equation}
The relevant terms in the Lagrangian are given by~\cite{Endo:2006ix}
\begin{equation}
	\mathcal L \;\simeq\;
%	 k\left[ \frac{1}{4}(\partial^2 X^\dagger) \tilde N_1\tilde N_1 + \frac{1}{2}X^\dagger (\partial_\mu \tilde N_1)(\partial^\mu \tilde N_1) - (\partial_\mu X^\dagger)(\tilde N_1 \partial^\mu \tilde N_1)
%	  \right]
k \left(
- (\partial^2 X^\dag) {\tilde N}_i {\tilde N}_i + m \phi N_i N_i 
\right)
	   + {\rm h.c.}
\end{equation}
The decay rate is then given by
\begin{equation}
		\Gamma(\Phi\to  N_i N_i)  
		\simeq \Gamma(\Phi\to\tilde N_i\tilde N_i) \simeq \frac{|k|^2}{16\pi}m^3.
\end{equation}
This results in  $T_{\rm R}$ comparable to (\ref{TR_N}) for $|k| \sim \mathcal O(1)$.
The advantage of this reheating process over the previous one is that we do not have 
to break the shift symmetry.

%%%%%%%%%%%%%%%%%%%%%%%%%%%%%%%%%%%%%
\subsubsection{Decay into the MSSM Higgs}
\label{sec:higgs}
%%%%%%%%%%%%%%%%%%%%%%%%%%%%%%%%%%%%%

Next let us consider  the inflaton decay into Higgs fields. We introduce the following coupling,
\begin{equation}
	W = \kappa X H_u H_d,
	\label{XHH}
\end{equation}
where $H_u$ and $H_d$ are MSSM up- and down-type Higgs doublets.
This explicitly breaks the $Z_2$ symmetry, and so, we assume $|\kappa| \ll 1$.\footnote{
In order to generate the $\mu$-term, we assume $H_uH_d$ is singlet under $Z_2$. 
In the case of $(2,4)$ model, the coupling (\ref{XHH}) does not break the $Z_2$ symmetry
as $X$ is neutral under $Z_2$. In this case, however, we need to assume that 
$d_0$ is sufficiently suppressed in order for the SUSY minimum to exist
in a region of $|{\rm Re}[\phi]| \lesssim 1$. 
} Similarly to the previous case, once the $Z_2$ breaking is introduced,
the inflaton acquires a non-zero VEV of order $|\kappa|/m$. Its effect on the reheating
is negligible if $|\kappa| \ll m$. 

We note that there is another subtle issue on the stability of the inflaton trajectory. 
By focusing on the $p=1$ term for the inflaton sector, $W = mX\phi$, the scalar potential reads
\begin{equation}
	V = |m\phi + \kappa H_u H_d|^2,
	% \supset \frac{i}{\sqrt{2}} m\varphi (\kappa H_uH_d - {\rm h.c.}).
\end{equation}
which shows that there is a flat direction along $m\phi + \kappa H_u H_d = 0$.
The presence of the flat direction may spoil the inflation, as it implies that 
the inflaton trajectory is unstable along the direction of $H_u H_d$. 
%$Thus, $H_u$ and $H_d$ should be stabilized at the origin for successful inflation.
Note that, in the absence of  the interaction with the inflaton, $H_u$ and $H_d$ can be 
stabilized at the origin by the Hubble-induced mass. For successful inflation, we need to keep 
both $H_u$ and $H_d$ stabilized
at the origin, which places an upper bound on $|\kappa|$,
%
%
%During inflation, $H_uH_d$ obtains a large mass squared of the order $\sim \kappa m \varphi$,
%which might drive the system into the SUSY vacuum spoiling the inflation.
%On the other hand, the following terms in the K\"ahler potential 
%%%
%\begin{equation}
%	K \supset |X|^2 (c_u |H_u|^2 + c_d |H_d|^2),
%\end{equation}
%%%
%with $c_u$ and $c_d$ being constants of $\mathcal O(1)$, generate SUSY breaking masses (squared) for $H_u$ and $H_d$
%of the order $\sim m^2 \varphi^2$ during inflation, which tend to stabilize Higgs fields at $H_u = H_d = 0$.
%Therefore, we need
%To this end, we need
%%
\begin{equation}
	|\kappa| \lesssim m \varphi_{\rm end} \sim \mathcal O(10^{-5}).
	\label{kappa}
\end{equation}
In the following  we assume $|\kappa| \ll 10^{-5}$ to avoid both too large inflaton VEV and
the instability of the inflaton trajectory.

%
%Once the condition (\ref{kappa}) is satisfied, the reheating occurs via the inflaton decay into the Higgs sector.
%After inflation, $\phi$ and $X$ are maximally mixed with each other due to the constant term $W_0$ in the superpotential
%(see Eq.~(\ref{W_gen})) and form the mass eigenstates as $\Phi_{\pm} = (\phi \mp X^\dagger)/\sqrt{2}$ with mass eigenvalues
%$m^2_{\pm} = m(m\pm m_{3/2})$~\cite{Kawasaki:2006gs}.

The interaction terms relevant for the reheating are
\begin{equation}
	\mathcal L = -(\kappa m \phi^\dagger H_u H_d + {\rm h.c.}) - (\kappa X \tilde H_u \tilde H_d + {\rm h.c.}).
\end{equation}
This induces the inflaton decay into the Higgs bosons and higgsinos.
The decay rate is given by
\begin{equation}
	\Gamma( \Phi \to H_uH_d) \simeq \Gamma(\Phi\to\tilde H_u\tilde H_d) = \frac{\kappa^2}{16\pi}m,
\end{equation}
and the reheating temperature is 
\begin{equation}
	T_{\rm R} \simeq 4\times 10^8\,{\rm GeV} 
	\left( \frac{\kappa}{10^{-6}} \right)
	\left( \frac{m}{10^{13}\,{\rm GeV}} \right)^{1/2}.
	\label{TR}
\end{equation}
Taking account of the constraint (\ref{kappa}), $T_{\rm R}$ cannot be higher than $\sim 10^{9}\,$GeV.
The thermal leptogenesis~\cite{Fukugita:1986hr} marginally works if $T_R$ is close to the upper bound. 
Otherwise we may need mild degeneracy between the right-handed neutrinos.

Another way to induce the inflaton decay into the Higgs sector is to introduce the following coupling:
\begin{equation}
	W = \kappa' \phi H_u H_d.
\end{equation}
As opposed to the case of $XH_uH_d$ coupling (\ref{XHH}), this does not affect the inflaton dynamics during inflation.
However, this coupling explicitly violates the shift symmetry.
Recalling that the order parameter of the shift symmetry breaking is $m\sim 10^{-5}$,
the coupling $\kappa'$ should also be of the order $\sim 10^{-5}$ or less. 
As a result, the reheating temperature is roughly same as (\ref{TR}).

Alternatively we may include the following interaction in the K\"ahler potential,
\beq
K \;=\; \kappa''(\phi+\phi^\dag) H_u H_d + {\rm h.c.},
\eeq
which respects the shift symmetry, but explicitly breaks the $Z_2$ symmetry. 
In order to avoid too large inflaton VEV, we require that the $Z_2$ breaking
is sufficiently suppressed, $|\kappa''| \ll 10^{-5}$. Then its contribution to the
reheating is negligible compared to the previous cases.

%%%%%%%%%%%%%%%%%%%%%%%%%%%%%%%%%%%%%
\subsection{Case without $Z_2$ symmetry}
%%%%%%%%%%%%%%%%%%%%%%%%%%%%%%%%%%%%%
Without the $Z_2$ symmetry, all the interactions in (\ref{W_gen}) are allowed.
Focusing on the two lowest order terms, the model is reduced to the $(1,2)$ model,
and the inflaton mass is fixed to be of order $10^{-5}$ by the Planck normalization 
of density perturbations.  As mentioned in Sec.~\ref{sec:pol}, we assume that
the $d_0$ term is sufficiently suppressed so that the inflaton potential has a SUSY
minimum in a region of $|{\rm Re}[\phi]| < 1$. This is the case if $|d_0| \lesssim 10^{-5}$
and $|c_1| \lesssim 1$. Since there is a priori no reason to expect $|c_1| \ll 1$, we
assume $|c_1| \sim 1$ and consider its contribution to the inflaton decay.

%%%%%%%%%%%%%%%%%%%%%%%%%%%%%%%%%%%%%
\subsubsection{Decay via the top Yukawa}
%%%%%%%%%%%%%%%%%%%%%%%%%%%%%%%%%%%%%

In the case without $Z_2$ symmetry, $\phi$ generically has an unsuppressed  linear term in the K\"ahler potential. 
The linear term can be also interpreted as the inflaton VEV, $\la \phi \ra = c_1$, as the VEV is induced 
if one shifts the origin of $\phi$ so that the linear term vanishes. 
As we have discussed at the beginning of this section, even without introducing couplings between the inflaton 
and MSSM sector by hand in the global SUSY limit,
the inflaton necessarily couples with the MSSM sector in SUGRA~\cite{Endo:2006qk}.
The decay mainly proceeds via the top Yukawa coupling and the decay rate is given by
\begin{equation}
	\Gamma(\Phi \to Q t H_u ) = \frac{3}{256\pi^3}|y_t|^2\left( \frac{\langle\phi\rangle}{M_P} \right)^2 \frac{m^3}{M_P^2},
\end{equation}
where $y_t$ is the top Yukawa coupling.
The reheating temperature is estimated to be
\begin{equation}
	T_{\rm R} \simeq 7\times 10^7\,{\rm GeV} |y_t|
	\left( \frac{\langle\phi\rangle}{10^{18}\,{\rm GeV}} \right)
	\left( \frac{m}{10^{13}\,{\rm GeV}} \right)^{3/2}.
\end{equation}

It is also possible to introduce the coupling like (\ref{XHH}).
In this case the reheating temperature is estimated as (\ref{TR}).

%%%%%%%%%%%%%%%%%%%%%%%%%%%%%%%%%%%%%
\subsubsection{Gravitino problem}  \label{sec:grav}
%%%%%%%%%%%%%%%%%%%%%%%%%%%%%%%%%%%%%

As extensively studied in Refs.~\cite{Kawasaki:2006gs,Asaka:2006bv,Dine:2006ii,Endo:2006tf,Endo:2007ih,Endo:2012yg}, the inflaton generally decays into the gravitino
which potentially leads to cosmological problems.
The gravitino abundance produced by the inflaton decay,
in terms of the number density-to-entropy-density ratio, is estimated as~\cite{Endo:2007ih}
\begin{equation}
	\frac{n_{3/2}}{s} \simeq 7\times 10^{-6} \left( \frac{10^9\,{\rm GeV}}{T_{\rm R}} \right)
	\left( \frac{\langle\phi\rangle}{10^{18}\,{\rm GeV}} \right)^2
	\left( \frac{m}{10^{13}\,{\rm GeV}} \right)^2,
\end{equation}
for $m < \Lambda$ where $\Lambda$ is the dynamical SUSY breaking scale, and
\begin{equation}
	\frac{n_{3/2}}{s} \simeq 9\times 10^{-8} \beta \left( \frac{10^9\,{\rm GeV}}{T_{\rm R}} \right)
	\left( \frac{\langle\phi\rangle}{10^{18}\,{\rm GeV}} \right)^2
	\left( \frac{m}{10^{13}\,{\rm GeV}} \right)^2,
\end{equation}
for $m > \Lambda$ with $\beta$ representing the model dependent parameter of order $1$-$10$.
To avoid the gravitino overproduction, we need one of the three options listed below.
\begin{itemize}
%\item The $Z_2$-symmetric model, e.g. $(p,q)=(1,3)$, leads to $\langle\phi\rangle = 0$ hence there is no gravitino overproduction,
%independently of the gravitino mass.
\item If the relation $m_z \ll m < \Lambda$ holds, where $m_z$ is the mass of the SUSY breaking field, 
the gravitino production rate is significantly suppressed, as explicitly considered in Ref.~\cite{Nakayama:2012hy}.
This solution matches with the gravitino mass $m_{3/2} = 100$--$1000$\,TeV.
\item If the gravitino is heavy enough to decay before the big-bang nucleosynthesis begins, and also the lightest SUSY particle (LSP)
decays via the small R-parity violation effects, there is no gravitino problem. 
This is possible for $m_{3/2} > \mathcal O(10)$\,TeV. In this case, however, we need another dark matter candidate, such as the axion.
\item If the gravitino is lighter than $\sim 10$\,eV, it thermalizes with the plasma and its relic abundance is so small that
it does not drastically affect the cosmological observations.
\end{itemize}

%%%%%%%%%%%%%%%%%%%%%%%%%%%%%%%%%%%%%
\section{Conclusions}  \label{sec:conc}
%%%%%%%%%%%%%%%%%%%%%%%%%%%%%%%%%%%%%

In this paper we have constructed the polynomial chaotic inflation model in the SUGRA framework,
and shown that they generically predict large enough tensor-to-scalar ratio to be detected in future B-mode experiments
while satisfying current observational constraints.
Therefore, non-detection of the B-mode at future experiments will exclude a large portion of the parameter space for the polynomial chaotic inflation.
From phenomenological points of view, the model with $Z_2$ symmetry is interesting since it avoids the gravitino overproduction from the inflaton decay.

We also found that the reheating temperature in such a model, Eq.~(\ref{TR}), is bounded from above so that the interaction (\ref{XHH}) does not disturb the inflaton dynamics.
The resulting upper bound reads $T_{\rm R} \lesssim 10^{10}\,$GeV.
Close this upper bound, the LSP produced by the thermally produced gravitino~\cite{Bolz:2000fu} can account for the present dark matter abundance, for the LSP mass of $\mathcal O(100)$\,GeV and the gravitino mass of $\mathcal O(100)$\,TeV.
This matches with the pure gravity mediation model~\cite{Ibe:2011aa},
where the gaugino masses are given by the anomaly-mediation effect~\cite{Giudice:1998xp}
while sfermions are as heavy as the gravitino and explains the 125\,GeV Higgs boson mass.

%%%%%%%%%%%%%%%%%%%%%%%%%%%%%%%%%%%%%%%%%%%%
\section*{Acknowledgments}
%%%%%%%%%%%%%%%%%%%%%%%%%%%%%%%%%%%%%%%%%%%%

This work was supported by the Grant-in-Aid for Scientific Research on
Innovative Areas (No. 21111006  [KN and FT],  No.23104008 [FT], No.24111702 [FT]),
Scientific Research (A) (No. 22244030 [KN and FT], 21244033 [FT], 22244021 [TTY]),  JSPS Grant-in-Aid for
Young Scientists (B) (No.24740135) [FT], and Inoue Foundation for Science [FT].  This work was also
supported by World Premier International Center Initiative (WPI Program), MEXT, Japan.

%%%%%%%%%%%%%%%%%%%%%%%%%%%%%%%%%%%%%

%%%%%%%%%%%%%%%%%%%%%%%%%%%%%%%%%%%%%


\begin{thebibliography}{99}
%%%%%%%%%%%%%%%%%%%%%%%%%%%%%%%%%%%%%


  %\cite{Guth:1980zm}
\bibitem{Guth:1980zm}
  A.~H.~Guth,
  %``The Inflationary Universe: A Possible Solution to the Horizon and Flatness Problems,''
  Phys.\ Rev.\  D {\bf 23}, 347-356 (1981);
  %\cite{Starobinsky:1980te}
%\bibitem{Starobinsky:1980te}
A.~A.~Starobinsky,
%``A New Type of Isotropic Cosmological Models without Singularity,''
Phys.\ Lett.\ B {\bf 91} (1980) 99;
%%CITATION = PHLTA,B91,99;%%
%\cite{Sato:1980yn}
%\bibitem{Sato:1980yn}
  K.~Sato,
  %``First Order Phase Transition of a Vacuum and Expansion of the Universe,''
  Mon.\ Not.\ Roy.\ Astron.\ Soc.\  {\bf 195}, 467-479 (1981).
  
  %\cite{Linde:1981mu}
\bibitem{Linde:1981mu}
A.~D.~Linde,
%``A New Inflationary Universe Scenario: a Possible Solution of the Horizon, Flatness, Homogeneity, Isotropy and Primordial Monopole Problems,''
Phys.\ Lett.\ B {\bf 108} (1982) 389;
%%CITATION = PHLTA,B108,389;%%
%\cite{Albrecht:1982wi}
%\bibitem{Albrecht:1982wi} 
  A.~Albrecht and P.~J.~Steinhardt,
  %``Cosmology for Grand Unified Theories with Radiatively Induced Symmetry Breaking,''
  Phys.\ Rev.\ Lett.\  {\bf 48}, 1220 (1982).
  %%CITATION = PRLTA,48,1220;%%
  
  %\cite{Ade:2013rta}
\bibitem{Ade:2013rta} 
  P.~A.~R.~Ade {\it et al.}  [ Planck Collaboration],
  %``Planck 2013 results. XXII. Constraints on inflation,''
  arXiv:1303.5082 [astro-ph.CO].
  %%CITATION = ARXIV:1303.5082;%%
  %5 citations counted in INSPIRE as of 26 Mar 2013
  
  

  

  %\cite{Martin:2013tda}
\bibitem{Martin:2013tda} 
  J.~Martin, C.~Ringeval and V.~Vennin,
  %``Encyclopaedia Inflationaris,''
  arXiv:1303.3787 [astro-ph.CO].
  %%CITATION = ARXIV:1303.3787;%%
  
    %\cite{Yamaguchi:2011kg}
\bibitem{Yamaguchi:2011kg} 
  M.~Yamaguchi,
  %``Supergravity based inflation models: a review,''
  Class.\ Quant.\ Grav.\  {\bf 28}, 103001 (2011)
  [arXiv:1101.2488 [astro-ph.CO]].
  %%CITATION = ARXIV:1101.2488;%%
  %14 citations counted in INSPIRE as of 08 May 2013
  
   %\cite{Linde:1983gd}
\bibitem{Linde:1983gd}
  A.~D.~Linde,
  %``Chaotic Inflation,''
  Phys.\ Lett.\ B {\bf 129}, 177 (1983).
  %%CITATION = PHLTA,B129,177;%%   

%\cite{Kawasaki:2000yn}
\bibitem{Kawasaki:2000yn} 
  M.~Kawasaki, M.~Yamaguchi and T.~Yanagida,
  %``Natural chaotic inflation in supergravity,''
  Phys.\ Rev.\ Lett.\  {\bf 85}, 3572 (2000)
  [hep-ph/0004243].
  %%CITATION = HEP-PH/0004243;%%
  %159 citations counted in INSPIRE as of 26 Mar 2013
  
  %\cite{Kawasaki:2000ws}
\bibitem{Kawasaki:2000ws} 
M.~Kawasaki, M.~Yamaguchi and T.~Yanagida,
  %``Natural chaotic inflation in supergravity and leptogenesis,''
  Phys.\ Rev.\ D {\bf 63}, 103514 (2001)
  [hep-ph/0011104].
  %%CITATION = HEP-PH/0011104;%%
  %58 citations counted in INSPIRE as of 26 Mar 2013
  
%\cite{Kallosh:2010ug}
\bibitem{Kallosh:2010ug} 
  R.~Kallosh and A.~Linde,
  %``New models of chaotic inflation in supergravity,''
  JCAP {\bf 1011}, 011 (2010)
  [arXiv:1008.3375 [hep-th]].
  %%CITATION = ARXIV:1008.3375;%%
  %44 citations counted in INSPIRE as of 28 Mar 2013
  
%\cite{Kallosh:2010xz}
\bibitem{Kallosh:2010xz} 
  R.~Kallosh, A.~Linde and T.~Rube,
  %``General inflaton potentials in supergravity,''
  Phys.\ Rev.\ D {\bf 83}, 043507 (2011)
  [arXiv:1011.5945 [hep-th]].
  %%CITATION = ARXIV:1011.5945;%%
  %23 citations counted in INSPIRE as of 28 Mar 2013

%\cite{Takahashi:2010ky}
\bibitem{Takahashi:2010ky} 
  F.~Takahashi,
  %``Linear Inflation from Running Kinetic Term in Supergravity,''
  Phys.\ Lett.\ B {\bf 693}, 140 (2010)
  [arXiv:1006.2801 [hep-ph]].
  %%CITATION = ARXIV:1006.2801;%%
  %18 citations counted in INSPIRE as of 28 Mar 2013
  
  %\cite{Nakayama:2010kt}
\bibitem{Nakayama:2010kt} 
  K.~Nakayama and F.~Takahashi,
  %``Running Kinetic Inflation,''
  JCAP {\bf 1011}, 009 (2010)
  [arXiv:1008.2956 [hep-ph]];
  %%CITATION = ARXIV:1008.2956;%%
  %21 citations counted in INSPIRE as of 28 Mar 2013
%\cite{Nakayama:2010sk}
%\bibitem{Nakayama:2010sk} 
%  K.~Nakayama, F.~Takahashi and ,
  %``Higgs Chaotic Inflation in Standard Model and NMSSM,''
  JCAP {\bf 1102}, 010 (2011)
  [arXiv:1008.4457 [hep-ph]];
  %%CITATION = ARXIV:1008.4457;%%
  %17 citations counted in INSPIRE as of 28 Mar 2013
%\cite{Nakayama:2010ga}
%\bibitem{Nakayama:2010ga} 
%  K.~Nakayama, F.~Takahashi and ,
  %``General Analysis of Inflation in the Jordan frame Supergravity,''
  JCAP {\bf 1011}, 039 (2010)
  [arXiv:1009.3399 [hep-ph]].
  %%CITATION = ARXIV:1009.3399;%%
  %8 citations counted in INSPIRE as of 28 Mar 2013

%\cite{Harigaya:2012pg}
\bibitem{Harigaya:2012pg} 
  K.~Harigaya, M.~Ibe, K.~Schmitz and T.~T.~Yanagida,
  %``Chaotic Inflation with a Fractional Power-Law Potential in Strongly Coupled Gauge Theories,''
  Phys.\ Lett.\ B {\bf 720}, 125 (2013)
  [arXiv:1211.6241 [hep-ph]].
  %%CITATION = ARXIV:1211.6241;%%
  
    %\cite{Nakayama:2013jka}
\bibitem{Nakayama:2013jka} 
  K.~Nakayama, F.~Takahashi and T.~T.~Yanagida,
  %``Polynomial Chaotic Inflation in the Planck Era,''
  arXiv:1303.7315 [hep-ph].
  %%CITATION = ARXIV:1303.7315;%%
  
    %\cite{Silverstein:2008sg}
\bibitem{Silverstein:2008sg} 
  E.~Silverstein, A.~Westphal,
  %``Monodromy in the CMB: Gravity Waves and String Inflation,''
  Phys.\ Rev.\ D {\bf 78}, 106003 (2008)
  [arXiv:0803.3085 [hep-th]].
  %%CITATION = ARXIV:0803.3085;%%
  %146 citations counted in INSPIRE as of 29 Mar 2013
    
  %\cite{McAllister:2008hb}
\bibitem{McAllister:2008hb} 
  L.~McAllister, E.~Silverstein, A.~Westphal,
  %``Gravity Waves and Linear Inflation from Axion Monodromy,''
  Phys.\ Rev.\ D {\bf 82}, 046003 (2010)
  [arXiv:0808.0706 [hep-th]].
  %%CITATION = ARXIV:0808.0706;%%
  %128 citations counted in INSPIRE as of 29 Mar 2013
  
    %\cite{Peiris:2013opa}
\bibitem{Peiris:2013opa} 
  H.~Peiris, R.~Easther, R.~Flauger,
  %``Constraining Monodromy Inflation,''
  arXiv:1303.2616 [astro-ph.CO].
  %%CITATION = ARXIV:1303.2616;%%
  %4 citations counted in INSPIRE as of 29 Mar 2013
  
    %\cite{Destri:2007pv}
\bibitem{Destri:2007pv} 
  C.~Destri, H.~J.~de Vega and N.~G.~Sanchez,
  %``MCMC analysis of WMAP3 and SDSS data points to broken symmetry inflaton potentials and provides a lower bound on the tensor to scalar ratio,''
  Phys.\ Rev.\ D {\bf 77}, 043509 (2008)
  [astro-ph/0703417].
  %%CITATION = ASTRO-PH/0703417;%%
  %34 citations counted in INSPIRE as of 01 Apr 2013
  
  %\cite{Linde:1983fq}
  \bibitem{Linde:1983fq}   
  A.~D.~Linde, 
   %``Supergravity And Inflationary Universe. (in Russian),''
  Pisma Zh.\ Eksp.\ Teor.\ Fiz.\  {\bf 37}, 606 (1983)
  [JETP Lett.\  {\bf 37}, 724 (1983)];
  %%CITATION = ZFPRA,37,606;%%
  %11 citations counted in INSPIRE as of 31 Mar 2013
%\cite{Linde:1984cd}
%\bibitem{Linde:1984cd}   
%A.~D.~Linde,  %``Primordial Inflation Without Primordial Monopoles,''
  Phys.\ Lett.\ B {\bf 132}, 317 (1983).
  %%CITATION = PHLTA,B132,317;%%
  %65 citations counted in INSPIRE as of 31 Mar 2013
    
    

    
    %\cite{Kallosh:2007wm}
\bibitem{Kallosh:2007wm} 
  R.~Kallosh and A.~D.~Linde,
  %``Testing String Theory with CMB,''
  JCAP {\bf 0704}, 017 (2007)
  [arXiv:0704.0647 [hep-th]].
  %%CITATION = ARXIV:0704.0647;%%
  %56 citations counted in INSPIRE as of 02 Apr 2013      

  
    %\cite{Croon:2013ana}
\bibitem{Croon:2013ana} 
  D.~Croon, J.~Ellis and N.~E.~Mavromatos,
  %``Wess-Zumino Inflation in Light of Planck,''
  arXiv:1303.6253 [astro-ph.CO].
  %%CITATION = ARXIV:1303.6253;%%


 %\cite{Cremmer:1983bf}
\bibitem{Cremmer:1983bf}
  E.~Cremmer, S.~Ferrara, C.~Kounnas and D.~V.~Nanopoulos,
  %``Naturally Vanishing Cosmological Constant in N=1 Supergravity,''
  Phys.\ Lett.\ B {\bf 133}, 61 (1983).
  %%CITATION = PHLTA,B133,61;%%

  %\cite{Ellis:1983sf}
\bibitem{Ellis:1983sf}
  J.~R.~Ellis, A.~B.~Lahanas, D.~V.~Nanopoulos and K.~Tamvakis,
  %``No-Scale Supersymmetric Standard Model,''
  Phys.\ Lett.\ B {\bf 134}, 429 (1984);
  %%CITATION = PHLTA,B134,429;%%
%\cite{Ellis:1984bm}
%\bibitem{Ellis:1984bm}
  J.~R.~Ellis, C.~Kounnas and D.~V.~Nanopoulos,
  %``No Scale Supersymmetric Guts,''
  Nucl.\ Phys.\ B {\bf 247}, 373 (1984);
  %%CITATION = NUPHA,B247,373;%%
  %\cite{Ellis:1983ei}
%\bibitem{Ellis:1983ei}
  J.~R.~Ellis, C.~Kounnas and D.~V.~Nanopoulos,
  %``Phenomenological SU(1,1) Supergravity,''
  Nucl.\ Phys.\ B {\bf 241}, 406 (1984).
  %%CITATION = NUPHA,B241,406;%%
  
  %\cite{Murayama:1993xu}
\bibitem{Murayama:1993xu} 
  H.~Murayama, H.~Suzuki, T.~Yanagida and J.~'i.~Yokoyama,
  %``Chaotic inflation and baryogenesis in supergravity,''
  Phys.\ Rev.\ D {\bf 50}, 2356 (1994)
  [hep-ph/9311326].
  %%CITATION = HEP-PH/9311326;%%
  %130 citations counted in INSPIRE as of 29 Mar 2013
  

%\cite{Ellis:2013xoa}
\bibitem{Ellis:2013xoa} 
  J.~Ellis, D.~V.~Nanopoulos and K.~A.~Olive,
  %``A No-Scale Supergravity Realization of the Starobinsky Model,''
  arXiv:1305.1247 [hep-th].
  %%CITATION = ARXIV:1305.1247;%%

  %\cite{Liddle:2000cg}
\bibitem{Liddle:2000cg} 
  A.~R.~Liddle and D.~H.~Lyth,
  ``Cosmological inflation and large scale structure,''
  Cambridge, UK: Univ. Pr. (2000).
  
\bibitem{Fukugita:1986hr}
  M.~Fukugita and T.~Yanagida,
  %``Baryogenesis Without Grand Unification,''
  Phys.\ Lett.\ B {\bf 174}, 45 (1986).

  %\cite{Endo:2006qk}
\bibitem{Endo:2006qk} 
  M.~Endo, M.~Kawasaki, F.~Takahashi and T.~T.~Yanagida,
  %``Inflaton decay through supergravity effects,''
  Phys.\ Lett.\ B {\bf 642}, 518 (2006)
  [hep-ph/0607170].
  %%CITATION = HEP-PH/0607170;%%
  
%\cite{Kawasaki:2006gs}
\bibitem{Kawasaki:2006gs} 
  M.~Kawasaki, F.~Takahashi and T.~T.~Yanagida,
  %``Gravitino overproduction in inflaton decay,''
  Phys.\ Lett.\ B {\bf 638}, 8 (2006)
  [hep-ph/0603265];
  %%CITATION = HEP-PH/0603265;%%
  %\cite{Kawasaki:2006hm}
%\bibitem{Kawasaki:2006hm} 
  %M.~Kawasaki, F.~Takahashi and T.~T.~Yanagida,
  %``The Gravitino-overproduction problem in inflationary universe,''
  Phys.\ Rev.\ D {\bf 74}, 043519 (2006)
  [hep-ph/0605297].
  %%CITATION = HEP-PH/0605297;%%
  
  %\cite{Lazarides:1991wu}
\bibitem{Lazarides:1991wu} 
  G.~Lazarides and Q.~Shafi,
  %``Origin of matter in the inflationary cosmology,''
  Phys.\ Lett.\ B {\bf 258}, 305 (1991).
  %%CITATION = PHLTA,B258,305;%%

%\cite{Asaka:1999yd}
\bibitem{Asaka:1999yd}
  T.~Asaka, K.~Hamaguchi, M.~Kawasaki, T.~Yanagida,
  %``Leptogenesis in inflaton decay,''
  Phys.\ Lett.\  {\bf B464}, 12-18 (1999)
  [hep-ph/9906366];
  %\cite{Asaka:1999jb}
%\bibitem{Asaka:1999jb}
  %T.~Asaka, K.~Hamaguchi, M.~Kawasaki, T.~Yanagida,
  %``Leptogenesis in inflationary Universe,''
  Phys.\ Rev.\  {\bf D61}, 083512 (2000)
  [hep-ph/9907559].
  
  %\cite{Endo:2006ix}
\bibitem{Endo:2006ix} 
  M.~Endo and F.~Takahashi,
  %``Non-thermal Production of Dark Matter from Late-Decaying Scalar Field at Intermediate Scale,''
  Phys.\ Rev.\ D {\bf 74}, 063502 (2006)
  [hep-ph/0606075].
  %%CITATION = HEP-PH/0606075;%%
  %36 citations counted in INSPIRE as of 19 May 2013
  
  %\cite{Asaka:2006bv}
\bibitem{Asaka:2006bv} 
  T.~Asaka, S.~Nakamura and M.~Yamaguchi,
  %``Gravitinos from heavy scalar decay,''
  Phys.\ Rev.\ D {\bf 74}, 023520 (2006)
  [hep-ph/0604132].
  %%CITATION = HEP-PH/0604132;%%
  %\cite{Dine:2006ii}
\bibitem{Dine:2006ii} 
  M.~Dine, R.~Kitano, A.~Morisse and Y.~Shirman,
  %``Moduli decays and gravitinos,''
  Phys.\ Rev.\ D {\bf 73}, 123518 (2006)
  [hep-ph/0604140].
  %%CITATION = HEP-PH/0604140;%%
%\cite{Endo:2006tf}
\bibitem{Endo:2006tf} 
  M.~Endo, K.~Hamaguchi and F.~Takahashi,
  %``Moduli/Inflaton Mixing with Supersymmetry Breaking Field,''
  Phys.\ Rev.\ D {\bf 74}, 023531 (2006)
  [hep-ph/0605091].
  %%CITATION = HEP-PH/0605091;%%
  
  %\cite{Endo:2007ih}
\bibitem{Endo:2007ih} 
  M.~Endo, F.~Takahashi and T.~T.~Yanagida,
  %``Anomaly-induced inflaton decay and gravitino-overproduction problem,''
  Phys.\ Lett.\ B {\bf 658}, 236 (2008)
  [hep-ph/0701042];
  %%CITATION = HEP-PH/0701042;%%
  %\cite{Endo:2007sz}
%\bibitem{Endo:2007sz} 
  %M.~Endo, F.~Takahashi and T.~T.~Yanagida,
  %``Inflaton Decay in Supergravity,''
  Phys.\ Rev.\ D {\bf 76}, 083509 (2007)
  [arXiv:0706.0986 [hep-ph]].
  %%CITATION = ARXIV:0706.0986;%%
  
  %\cite{Endo:2012yg}
\bibitem{Endo:2012yg} 
  M.~Endo, K.~Hamaguchi and T.~Terada,
  %``Scalar Decay into Gravitinos in the Presence of D-term SUSY Breaking,''
  Phys.\ Rev.\ D {\bf 86}, 083543 (2012)
  [arXiv:1208.4432 [hep-ph]].
  %%CITATION = ARXIV:1208.4432;%%

%\cite{Nakayama:2012hy}
\bibitem{Nakayama:2012hy} 
  K.~Nakayama, F.~Takahashi and T.~T.~Yanagida,
  %``Eluding the Gravitino Overproduction in Inflaton Decay,''
  Phys.\ Lett.\ B {\bf 718}, 526 (2012)
  [arXiv:1209.2583 [hep-ph]].
  %%CITATION = ARXIV:1209.2583;%%

%\cite{Bolz:2000fu}
\bibitem{Bolz:2000fu} 
  M.~Bolz, A.~Brandenburg and W.~Buchmuller,
  %``Thermal production of gravitinos,''
  Nucl.\ Phys.\ B {\bf 606}, 518 (2001)
  [Erratum-ibid.\ B {\bf 790}, 336 (2008)]
  [hep-ph/0012052];
  %%CITATION = HEP-PH/0012052;%%
  %\cite{Pradler:2006qh}
%\bibitem{Pradler:2006qh} 
  J.~Pradler and F.~D.~Steffen,
  %``Thermal gravitino production and collider tests of leptogenesis,''
  Phys.\ Rev.\ D {\bf 75}, 023509 (2007)
  [hep-ph/0608344];
  %%CITATION = HEP-PH/0608344;%%
  %\cite{Pradler:2006hh}
%\bibitem{Pradler:2006hh} 
  %J.~Pradler and F.~D.~Steffen,
  %``Constraints on the Reheating Temperature in Gravitino Dark Matter Scenarios,''
  Phys.\ Lett.\ B {\bf 648}, 224 (2007)
  [hep-ph/0612291];
  %%CITATION = HEP-PH/0612291;%%
  %\cite{Rychkov:2007uq}
%\bibitem{Rychkov:2007uq} 
  V.~S.~Rychkov and A.~Strumia,
  %``Thermal production of gravitinos,''
  Phys.\ Rev.\ D {\bf 75}, 075011 (2007)
  [hep-ph/0701104].
  %%CITATION = HEP-PH/0701104;%%
  
  %\cite{Ibe:2011aa}
\bibitem{Ibe:2011aa} 
  M.~Ibe and T.~T.~Yanagida,
  %``The Lightest Higgs Boson Mass in Pure Gravity Mediation Model,''
  Phys.\ Lett.\ B {\bf 709}, 374 (2012)
  [arXiv:1112.2462 [hep-ph]];
  %%CITATION = ARXIV:1112.2462;%%
%\cite{Ibe:2012hu}
%\bibitem{Ibe:2012hu} 
  M.~Ibe, S.~Matsumoto and T.~T.~Yanagida,
  %``Pure Gravity Mediation with m_{3/2} = 10-100TeV,''
  Phys.\ Rev.\ D {\bf 85}, 095011 (2012)
  [arXiv:1202.2253 [hep-ph]].
  %%CITATION = ARXIV:1202.2253;%%

%\cite{Giudice:1998xp}
\bibitem{Giudice:1998xp} 
  G.~F.~Giudice, M.~A.~Luty, H.~Murayama and R.~Rattazzi,
  %``Gaugino mass without singlets,''
  JHEP {\bf 9812}, 027 (1998)
  [hep-ph/9810442];
  %%CITATION = HEP-PH/9810442;%%
  %\cite{Randall:1998uk}
%\bibitem{Randall:1998uk} 
  L.~Randall and R.~Sundrum,
  %``Out of this world supersymmetry breaking,''
  Nucl.\ Phys.\ B {\bf 557}, 79 (1999)
  [hep-th/9810155].
  %%CITATION = HEP-TH/9810155;%%


%%%%%%%%%%%%%%%%%%%%%%%%%%%%%%%%%%%%%
\end{thebibliography}
\end{document}